# Disconnection-mediated twin embryo growth in Mg

Yang Hu [a], Vladyslav Turlo [b], Irene J. Beyerlein [c,d], Subhash Mahajan [e], Enrique J. Lavernia [a], Julie M. Schoenung [a], Timothy J. Rupert [a,b,*]

[a] Department of Materials Science and Engineering, University of California, Irvine, CA 92697, USA
[b] Department of Mechanical and Aerospace Engineering, University of California, Irvine, CA 92697, USA
[c] Mechanical Engineering Department, University of California, Santa Barbara, CA 93106, USA
[d] Materials Department, University of California, Santa Barbara, CA 93106, USA
[e] Department of Materials Science and Engineering, University of California, Davis, CA 95616, USA
* To whom correspondence should be addressed: trupert@uci.edu

**Abstract**

While deformation twinning in hexagonal close-packed metals has been widely studied due to its substantial impact on mechanical properties, an understanding of the detailed atomic processes associated with twin embryo growth is still lacking. Conducting molecular dynamics simulations on Mg, we show that the propagation of twinning disconnections emitted by basal-prismatic interfaces controls the twin boundary motion and is the rate-limiting mechanism during the initial growth of the twin embryo. The time needed for disconnection propagation is related to the distance between the twin tips, with widely spaced twin tips requiring more time for a unit twin boundary migration event to be completed. Thus, a phenomenological model, which unifies the two processes of disconnection and twin tip propagation, is proposed here to provide a quantitative analysis of twin embryo growth. The model fits the simulation data well, with two key parameters (twin tip velocity and twinning disconnection velocity) being extracted. In addition, a linear relationship between the ratio of twinning disconnection velocity to twin tip velocity and the applied shear stress is observed. Using an example of twin growth in a nanoscale single crystal from the recent literature, we find that our molecular dynamics simulations and analytical model are in good agreement with experimental data.

# 1. Introduction

Mg is the lightest structural material and has a high specific strength, which makes it promising for applications in the aerospace and automotive industries. The use of Mg and Mg-rich alloys in these applications would significantly reduce fuel costs and carbon dioxide emissions, thus contributing to lower environmental impact [1, 2]. However, Mg and Mg-rich alloys tend to be brittle due to the limited number of easily-activated slip systems in hexagonal close-packed (hcp) materials [3, 4]. While basal slip can be activated at less than 1 MPa of resolved shear stress [5, 6], pyramidal slip requires resolved shear stresses as high as ~40 MPa to operate [5]. Unlike face-centered cubic and body-centered cubic metals, which have a plethora of easily-activated slip systems, allowing for arbitrary morphology changes during plasticity, basal slip alone cannot support arbitrary plastic strains, particularly those aligned along the c-axis of the crystal. In contrast, deformation along the c-axis can be accommodated through twinning, meaning this is an important deformation mode contributing to the plasticity of Mg.

Common twinning modes in Mg include tension twins in both $\{10\bar{1}2\}\langle\bar{1}011\rangle$ and $\{11\bar{2}1\}\langle\bar{1}\bar{1}26\rangle$ and contraction twins in both $\{10\bar{1}1\}\langle10\bar{1}\bar{2}\rangle$ and $\{11\bar{2}2\}\langle11\bar{2}\bar{3}\rangle$ [7]. Here we focus on the $\{10\bar{1}2\}$ tension twin, as it is favored for stretching the c-axis and is the most profuse twinning mode in hcp metals [8, 9]. $\{10\bar{1}2\}$ twins are easier to nucleate due to the lower shear and shuffle displacements required of atoms in order to twin along the $\{10\bar{1}2\}$ plane [10]. In addition, $\{10\bar{1}2\}$ twinning planes can interact with lattice defects, such as dislocations, to form interfacial twinning disconnections, dislocations with a step character, that have been proposed as being responsible for twin boundary (TB) motion [11-13]. Therefore, $\{10\bar{1}2\}$ twins can easily grow as well. Understanding the growth mechanisms of $\{10\bar{1}2\}$ deformation twins is essential for explaining plasticity in hcp metals, and will be critical in future attempts to further enhance the ductility of Mg and Mg-rich alloys.

The formation of mature twins with the length and thickness of several hundred nanometers or micrometers starts from the nucleation and growth of twin embryos. Studies of twin embryo nucleation can be traced back to the 1950s when both homogeneous [14] and heterogeneous nucleation mechanisms [15] were first proposed. Heterogeneous nucleation assisted by slip dislocations, grain boundaries, free surfaces, or other defects requires less stress and energy, making it the likely dominant mechanism. In support of the concept of heterogeneous nucleation, Mendelson later found that the nonplanar dissociation of $<a>$, $<a+c>$, and $<c>$ dislocations can all generate glissile twinning dislocations [16, 17]. More recent work by Wang et al. [18] showed that $\{10\bar{1}2\}$ tension twins can be nucleated through partial dislocation dipoles with Burgers vectors parallel to the twinning direction, forming twin embryos bounded by TBs and basal-prismatic (BP) or prismatic-basal (PB) interfaces. BP/PB interfaces are asymmetric boundaries attached to TBs with a basal lattice facing a prismatic lattice (the choice of BP or PB is determined by whether the basal lattice plane is in the twin or in the matrix) [19-21]. In contrast to prior work on twin embryo nucleation, twin embryo growth mechanisms have been largely neglected due to the difficulties of tracking their growth process experimentally. Twin embryos typically grow rapidly and become mature twins by the time they are observed. However, the early stages of the twin embryo growth process should dramatically impact the morphology of mature twins, underscoring the critical need for a comprehensive model of twin embryo growth.

The growth of a twin embryo requires a combination of growth in two critical directions: (1) twin thickening through TB migration and (2) lateral motion of the twin tip (TT), which can also be called TT propagation. Twin thickening involves the nucleation and propagation of twinning disconnections (or twinning dislocations) on existing twin planes [22-27]. Twinning disconnections are defined as interfacial dislocations with both a Burgers vector and a step height, which exist to maintain the coherency at TBs during their migration. Using nudged elastic band calculations, a higher activation energy of homogeneous disconnection nucleation than propagation was reported [28, 29], suggesting that disconnection nucleation can be the rate-limiting process of TB motion. Twinning disconnections may also nucleate heterogeneously, such as when the release of twinning disconnections from BP/PB interfaces was observed by El Kadiri et al. [9]. During the early stages of twin growth, BP/PB interfaces make up a high fraction of the overall interfacial area separating the twin embryo from the matrix [18, 22, 30] and their presence makes it unclear which process controls twin embryo growth: nucleation or propagation of twinning disconnections. On the other hand, it was demonstrated both experimentally [19-21] and with molecular dynamics (MD) simulations [18, 30] that BP/PB interfaces are present at the TTs and may even drive their motion by disconnection formation and glide on such interfaces. However, the few existing reports on the lateral motion of TTs focus only on revealing the microstructure of the twin tip, rather than studying the migration process and growth mechanism [22, 30-33].

To shed light on the early stages of twin growth, we first conducted MD simulations of twin embryo growth in a pure Mg single crystal under different applied shear strains. Twinning disconnections were formed at the intersections of TBs with BP/PB interfaces as the result of nucleation and glide of disconnections on BP/PB interfaces. The later propagation of twinning disconnections along the TBs then drives the TB migration. We also discovered in our MD simulations that the TB velocity decreases as the twin length increases, suggesting that TB motion is limited by disconnection propagation. To extend our MD findings to much larger length- and time-scales and to establish a connection with experimental observations, we developed a phenomenological model for twin embryo growth that is driven by propagating disconnections. This model is used to obtain analytical expressions for the evolution of twin length and thickness as a function of time. We demonstrated that the morphology of a twin embryo at any moment in time is determined by the ratio of the twinning disconnection velocity to the TT velocity, which was found to be proportional to the applied shear strain. The phenomenological model was tested against data extracted from the experimental work of Yu et al. [34] on twinning in Mg nanoscale single crystals, confirming that our model is able to describe the early stages of twinning deformation.

## 2. Computational Methods

To study the atomic scale mechanisms responsible for twin embryo growth, we first perform MD simulations using the Large-scale Atomic/Molecular Massively Parallel Simulator (LAMMPS) package [35] and a Modified Embedded-Atom Method (MEAM) potential developed by Wu et al. that was constructed to accurately model the plastic deformation and fracture of Mg [36]. Atoms are relaxed under an NVT ensemble at 1 K or 300 K using Nose-Hoover thermostat, with the temperature adjusted every 500 time steps with one integration step of 0.1 fs. The low temperature simulations at 1 K are used to inhibit any effect of thermal fluctuations on twin growth, making it easier to identify the key deformation mechanisms. The simulations conducted at 300 K are used to compare with the experimental results on twinning in nanoscale Mg single crystals [34]. Structural analysis and visualization of atomic configurations are performed using the open-source visualization tool OVITO [37]. The Polyhedral Template Matching (PTM) method [38] is used to characterize the local crystalline structure and orientation associated with each atom in the system. Mg atoms with an hcp structural environment are colored blue, while those with any other structural environments are colored white.

A schematic of the simulation cell is shown in Fig. 1, with the X-axis along the $[1\bar{2}10]$ direction, the Y-axis along the twinning direction $[10\bar{1}1]$, and the Z-axis perpendicular to the $(\bar{1}012)$ plane (approximately parallel to the $[10\overline{1m}]$ direction with m = 1.139, as reported by Ostapovets and Gröger [23]). The dimensions of the simulation box are $5.15 \times 76.8 \times 55$ nm$^3$ in the X, Y, and Z directions, respectively, and the sample contains ~950,000 atoms. The dimensions along the Y and Z directions are chosen to minimize interactions between the twin embryo and the fixed-atom region, as well as interactions between the twin embryo and its periodic image. One twin embryo with a length of 7.0 nm ($l_0$, the initial length of the twin embryo) and a thickness of 4.3 nm ($h_0$, the initial thickness of the twin embryo) is inserted at the center of a simulation box containing the perfect Mg crystal lattice, using the Eshelby method reported by Xu et al. [30]. The ratio of the initial embryo length to the initial embryo thickness is 1.6. To test the effect of initial shape and size on the twin embryo growth process, the simulations are repeated using two other initial twin embryo sizes with different aspect ratios: (1) length $l_0 = 8.2$ nm and thickness $h_0 = 7.5$ nm ($l_0/h_0 \approx 1.1$), and the other with $l_0 = 15.4$ nm and $h_0 = 4.1$ nm ($l_0/h_0 \approx 3.7$).

The following procedure for twin embryo insertion was used. First, a piece of the initial Mg lattice bound by two conjugate twinning planes in the $\{10\bar{1}2\}$ plane family is extracted (the horizontal twinning plane is the $(\bar{1}012)$ plane and the one almost perpendicular to it is the $(10\bar{1}2)$ plane). Then a twinned region is obtained by applying mirror symmetry to the extracted region, which is subsequently introduced into the lattice. The shape of the twinned region is slightly different from the shape of the extracted region of the initial lattice. In order to match the shape of these two regions, a shear is applied to the twinned region. Finally, the sample is structurally relaxed using energy minimization. If there are no shear strains applied to the sample, the twin embryo is not stable and will shrink until it disappears due to the local stress profile. Therefore, shear strains are applied parallel to the $(\bar{1}012)$ planes in the $[10\bar{1}1]$-direction to stabilize the twin embryo. The minimum shear strain which can stabilize the twin embryo is around 5%. For the following shear simulation, the whole simulation box is first shear back to a non-strain state, and then deformed at certain strain level. The twin embryo obtained after energy minimization is bound by a series of connected TBs, conjugate TBs, and BP/PB interfaces. Specifically, the upper and lower plane is the $(\bar{1}012)$ twinning plane (referred to simply as "TBs" from now on, as these defects would be the only TBs remaining once the twin embryo has matured), while the vertical planes are the $(10\bar{1}2)$ twinning planes (these are the twin tips, so they are referred to simply as "TTs" from now on). The two twinning planes are connected by BP and PB interfaces. Various morphologies of TTs of the $\{10\bar{1}2\}$ twins have been reported in prior literature [31, 32, 39, 40]. The TTs can either be comprised of BP/PB interfaces only [31, 32], or be made of planes which are perpendicular to primary twinning planes, such as planes contain interfacial dislocations and are separated by BP/PB interfaces [39, 40]. In our simulations, the initial TTs are defect free.

Periodic boundary conditions are applied in the X- and Y-directions, while the top and bottom surfaces in the Z-direction are free surfaces. Shear strains are applied uniformly by displacing all atoms in the simulation box. To maintain a constant shear strain, atoms within a 2 nm thick band at the top and bottom of the box are fixed, while the rest of atoms are free to move. The thickness of the fixed region is chosen to be above the cutoff distance of the interatomic potential to ensure that the evolving atoms see the fixed strain condition, not the free surface on top and bottom. The motivation behind this type of loading is to provide a constant driving force for twin embryo growth, by putting the matrix into a state where shear stress is available to drive growth. The expected driving force (the matrix stress) can be obtained by finding the shear stress in a single crystal with the matrix configuration, as shown in Fig. S1. The overall or global shear stress in the combined matrix-twin embryo sample will decrease as the system evolves and the twin fraction grows, but we will show that the assumption of a constant matrix stress is reasonable for the conditions studied here. More details will be provided in the Results section of this paper. Fig. 2 shows the distribution of atomic shear stress in snapshots taken at 10 ps and 25 ps, for samples deformed at 5%, 7%, and 11% shear strain. The mobile-atom region and fixed atom region are separated by dashed black lines, while the boundary of the twin embryo is demarcated by a solid black line. The distribution of atomic shear stress does not change for samples deformed at 5% shear strain, meaning the twin embryo is stable. In contrast, the samples deformed at 7% and 11% shear strain evolve and the twins grow with time. For the 11% shear strain-deformed sample, the position of the upper and lower TB is too close to the fixed-atom region and a strong interaction between them might be expected. The stress fields associated with the TTs also begin to interact with one another across the periodic boundary conditions. Thus, we focus on shear strains ranging from 6% to 10%, so that the lowest shear strain can activate twin embryo growth during MD timescales, yet the highest shear strain would not lead to rampant twin embryo growth such that there is an interaction with the fixed-atom regions and its periodic images.

To quantify the growth of the twin embryo, its length is calculated as the difference between the positions of the two TTs, while its thickness is calculated as the difference between the positions of the upper and lower TBs. To determine the positions of the TBs and TTs at any given moment, the atoms in the twinned region should be correctly identified first via the orientation information of each atom. The PTM method implemented in OVITO gives a point-to-point correspondence between the template structures and the structures to be analyzed, by minimizing the root-mean-square deviation of atom positions from the templates and simulated structures [38]. The lattice orientation information can be encoded as an orientation quaternion, $q = q_w + q_x i + q_y j + q_z k$. Fig. 3(a) shows that component $q_w$ is the best to distinguish the twin embryo from the matrix, using a threshold value of $q_w = 0.8$. This criterion can be used for samples deformed at different strains and configurations taken at different times, as shown in Fig. 3(b). To identify each boundary position, the simulation box is divided into 100 bins along the Y or Z-axis (displayed in Fig. 4), and then the number of atoms with $q_w > 0.8$ is counted in each bin and plotted versus the position of the bin. The peak in Fig. 4(c) then represents the twin embryo. Since the number of atoms with $q_w > 0.8$ in each bin changes with time as the twin grows, these values are normalized by the peak value in order to set up a consistent criterion for twin identification. Finally, the full width at half maximum values along the Y or Z-axis represent the positions of the TBs or TTs, respectively.

## 3. Results

Atomistic simulations at 1 K were first used to reveal the atomic scale mechanisms responsible for the expansion of a twin embryo under different applied shear strains. For example, Fig. 5(a) shows the boundary atoms of the twin embryo in the sample deformed under 7% shear strain, while Fig. 5(b) shows the corresponding boundary positions extracted at different moments in time. The propagation of TTs is found to be faster than twin thickening, with a final twin length that is ~2.3 times larger than the twin thickness at the end of the simulation (32 ps). Fig. 5(b) shows a linear relationship between the TT position and time, which suggests the TTs move at a constant velocity. In contrast, TB motion slows down over time, indicating a decrease in the TB velocity.

To explore the driving force for twin embryo growth, the global shear stress that acts parallel to the Y-direction and, therefore, to the top TB, is plotted as a function of time. The applied global, $\tau_{Global}$, shear stress includes the contribution from the matrix and the twin and can be expressed as:

$$\tau_{Global} = (\tau_{Matrix} \times (1 - f_{twin})) + (\tau_{Twin} \times f_{twin}) \tag{1}$$

where $\tau_{Matrix}$ is the shear stress in the matrix with no twin, measured in single crystals deformed at 6-10% shear strain and shown in Fig. S1. The $\tau_{Twin}$ is then obtained by fitting Eqn. (1) to the variation of global shear stress with the twin volume fraction. These values are negative, consistent with prior experimental results [41]. The $f_{twin}$ is the twin volume fraction. As an example, we consider the sample in Fig. 6(a), where a decreasing trend is found when a shear strain of 7% is applied. Fig. 6(b) shows the global shear stress as a function of twin volume fraction for different strains, confirming the linear relationship between these two variables. As the applied shear strain increases, both $\tau_{Matrix}$ and $\tau_{Twin}$ increase, while their difference (corresponding to the slope of the data in Fig. 6(b)) stays nearly the same. The twin volume fraction increases with time, as shown in Fig. S2 of the Supplementary Material.

The normal stresses acting parallel to the PB and BP interfaces are also presented in Figs. 6(c) and (d), respectively. The applied shear strain causes a region under compression near the PB interface and a region under tension near the BP interface. This means that in the vicinity of the PB interface, the basal lattice in the matrix is compressed more than the prismatic lattice in the twin embryo. Near the BP interface, the prismatic lattice in the matrix is stretched more than the basal lattice in the twin embryo. Basic hcp crystallography shows that the repeating unit of the basal lattice has a slightly larger lattice constant ($\sqrt{3}a$, where $a$ is the $\langle 11\bar{2}0 \rangle$ lattice parameter of Mg, with a value of 0.320 nm [42]) than the repeating unit of the prismatic lattice ($c$, where $c$ is the $\langle 0001 \rangle$ lattice parameter of Mg, with a value of 0.520 nm [42]). Compression of the basal lattice makes the transformation from basal to prismatic lattice easier and induces the expansion of the prismatic lattice. In contrast, tension of the prismatic lattice makes the transformation from prismatic to basal lattice easier, corresponding to the growth of the basal lattice. As a whole, the stress field around the BP/PB interfaces drives the outward motion of these interfaces.

Our MD simulations also show that most twinning disconnections form at the intersections of the TBs and BP/PB interfaces, and then propagate laterally to enable the growth of the twin embryo. An example is displayed in Fig. 7, where only the boundary atoms (white atoms) are shown and the hcp atoms in the matrix and twin are represented by the light blue background. In the first frame (8 ps) of Fig. 7(a), a twinning disconnection forms at the intersection on the right, followed by the formation of a second twinning disconnection at the intersection on the left in the second frame (9 ps). These two twinning disconnections move laterally towards each other in the second and third frames, similar to observations of disconnection migration on twinning planes by Xu et al. [30], Ostapovets and Gröger [23], and El Kadiri et al. [9]. Because of this disconnection propagation, the upper TB eventually moves upward one step height. At the same time, the TTs migrate laterally as well, as shown by comparing the TT positions with the dashed black line denoting the position at 8 ps. The spatial distribution of atoms near the first twinning disconnection at 8 ps and 11 ps is displayed in Fig. 7(b), with pairs of atoms near each other denoting atoms at the two different times and the displacements between these atoms indicated by white arrows. The atoms, which were once in the matrix (red atoms) but joined the twin (green atoms) later, exhibit displacements in different directions, although there is a pattern with a repeating periodicity. Wang et al. [22] reported the movements of atoms in Zn samples containing even-layer-thick twins or odd-layer-thick twins of different thicknesses. Comparing their work with ours, it shows that the atomic displacements are not the same, although a shear strain parallel to the $\{10\bar{1}2\}$ plane is applied in both cases. However, only TBs exist in their simulations, while we have TBs, TTs, and PB/BP interfaces in our simulated twin embryos. Fig. 7(b) shows that most atomic displacements are a combination of horizontal displacements of atoms parallel to the twin planes and vertical displacements parallel to the TTs. A zoomed-in view of the atomic structure of a twinning disconnection is displayed in Fig. 7(c) with the Burgers vector and step character shown. The Burgers vector of a twinning disconnection is $b_t = \frac{1}{15}[10\bar{1}1]$ and the step height, $h_d$, is 0.38 nm. In Fig. 7(d), the distances traveled by several twinning disconnections, analyzed during the twin embryo growth, are plotted as a function of simulation time, showing linear trends for each twinning disconnection. The slopes of the dashed lines are very similar, indicating a relatively constant velocity of the twinning disconnections for a given applied shear strain during the time period considered.

To further investigate the propagation of TTs, a sequence of atomic snapshots of the right TT is shown in Fig. 8(a), with magnified views of the PB interface in Fig. 8(b). As the TT moves to the right, several disconnections with step heights of one interplanar spacing are observed on the PB interface and are marked by black arrows in Fig. 8(b). These appear to be edge disconnections (disconnections with only edge Burgers vector character) with step heights of one interplanar spacing (*c*/2) on BP/PB interfaces, the same as those reported in the work of Zu et al. [43]. These disconnections move towards the intersections with the TB or the TT, causing the PB interfaces to migrate in a diagonal direction to the upper right of the simulation cell. At the same time, twinning disconnections start to form at the intersections but on twin planes, as denoted in the third and fourth frames of Fig. 8(b) by yellow arrows. One twinning disconnection is observed at the top intersection after the PB interface propagates two atomic planes in the diagonal direction, indicating that two of the disconnections on the PB interface contribute to the formation of one twinning disconnection. Details of the atomic structures of disconnections on BP/PB interfaces in our work can be found in Fig. 9. At 6.1 ps, there is a one-layer disconnection formed on the BP interface. The disconnection front involves atoms from two different BP planes, and these atoms are also in different planes along the $[\bar{1}010]$-direction, so part of atoms in lower BP plane are covered as shown in the view from the $[\bar{1}010]$-direction. At 10.15 ps, there are two one-layer disconnections formed on the PB interface, and one of them is on the top of the other one. At 10.5 ps, there is a two-layer disconnection formed on the BP interface. Barrett and El Kadiri reported the generation of disconnections with step heights of two interplanar spacings (*c*) on BP/PB interfaces by two disconnections with step heights of one interplanar spacing, and a final transformation to twinning disconnections with the involvement of disclinations (line defects where rotational symmetry is violated) at the intersections of TBs and BP/PB interfaces [44]. A similar transformation between disconnections on BP/PB interfaces and twinning disconnections was also found by Sun et al. [31], and Zu et al. [43]. According to our observations, PB interfaces move the fastest among all the boundaries of the embryo; therefore, they act as nucleation sites for twinning disconnections.

Formation and gliding of disconnections on TTs is also observed. These disconnections are also two-layer disconnections with Burgers vector parallel to the twinning direction of the $(10\bar{1}2)$ plane. Fig. 10(a) shows the atomic structure of one twinning disconnection formed on the TT, while Fig. 10(b) shows the Burgers vector of this disconnection, determined to be $b_t = \frac{1}{15}[\bar{1}011]$. In the studies by Braisaz et al. [40] and Lay et al. [39], interfacial dislocations with Burgers vectors approximately equal to four twinning dislocations and seven twinning dislocations were reported. In our simulations, such large interfacial dislocations are not found. Unlike the twinning disconnections on TBs that have relatively straight disconnection lines, the twinning disconnections on TTs have much more tortuous disconnection lines. Since the Burgers vectors of twinning disconnections are parallel to the twinning direction, tortuous disconnection lines indicate mixed characters of the twinning disconnections on TTs. A comparison of the twinning disconnections on TBs and TTs is made in Fig. 8(c) using the normal views of boundary atoms on the two different TBs (XY plane for the TB and YZ plane for the TT). The atoms are colored by their Y or Z positions. For instance, in the normal view of the TT, the dark blue atoms have the smallest Y positions and encompass the initial TT, while the red atoms have the largest Y positions and cover the new TT, which has moved one-step height. The yellow atoms are located between these two atomic planes and delineate the disconnection lines associated with the twinning disconnections. In the normal view of the TB, there are three twinning disconnections. The two furthest to the right have not formed on the same twinning plane, so their disconnection lines lie at different Z heights and appear as different colors. It can be clearly seen that almost all the white atoms are aligned. Although some tortuosity of the disconnection lines is also observed on the TBs (yellow atoms), this effect is much more prominent on the TT. Schematics of the TB and TT are also shown with the disconnection lines of twinning disconnections denoted on the two planes. Since the Burgers vectors of twinning disconnections on both planes are parallel to the twinning direction, tortuous disconnection lines indicate twinning disconnections with mixed character. Transformations from the disconnections on BP/PB interfaces to the twinning disconnections on TTs are also observed. Such transformations are similar to those from disconnections on BP/PB interfaces to twinning disconnections on TBs. In the last frame of Fig. 8(b), the twinning disconnection at the intersection of the PB interface and the TT is formed in this way. Fig. 11 shows an example of a transformation in detail, with an enlarged view of the BP interface at the top left corner presented. The two one-layer disconnections on the BP plane which contribute to the formation of twinning disconnection near the upper intersection between the BP interface and TT are D1 and D2. The new twinning disconnection is formed on TT2. D1 is bound by the black and red BP planes, while D2 is bound by the red and dark blue BP planes. At 5.8 ps, D1 is already formed, while D2 starts to form on BP2. At 6.4 ps, D1 and D2 transform into a two-layer disconnection. At 6.65 ps, one new twinning disconnection starts to form on TT2. Two other disconnections observed in this figure, D3 and D4, contribute to the formation of subsequent twinning disconnections. Homogeneous nucleation of the twinning disconnection on the TTs is often observed (see Fig. S3 and Video 1 in the Supplementary Material), which is likely due to the larger local shear stress in front of TTs than the shear stress in front of TBs. In Fig. 2 and Fig. S4, the local shear stress distribution of the sample that is deformed to 7% shear strain is shown. There are two blue regions next to the TBs on the matrix side, indicating regions with lower shear stress, which is not favorable for disconnection nucleation. The different types of twinning disconnections formed on the TB and TT lead to different migration behavior of the TB and the TT. Schematics showing the propagation of twinning disconnections on the TT and the TB can be found in Fig. S5. On TBs, the disconnection fronts do not move as a whole, but instead part of the disconnection front pushes forward first. This is followed by motion along the X-direction in a rapid, un-zipping manner. Consequently, the entire disconnection front has moved forward one step. The motion of the TBs upward therefore occurs when a twinning disconnection has traveled across the entire TB to the opposite PB/BP interface. Therefore, TB motion is rate-limited by disconnection propagation. In contrast, a larger number of disconnection loops are found to nucleate homogeneously on the TTs. These features can merge with twinning disconnections coming from the BP/PB interfaces, or other homogeneously nucleated twinning disconnections even before their X-direction propagation is finished. Unlike the TBs where multiple disconnections at different heights can be found (see Figs. 5(a) and 8(c)), no accumulation of disconnections is observed on the TTs, suggesting that the motion is nucleation-controlled, rather than propagation-controlled. In addition, Fig. 5(b) shows the TT position as a function of time, where the linear shape shows that TT motion does not slow with growth of the twin embryo, as would be expected if propagation was the controlling mechanism.

**4.     Discussion**

Our MD simulation results capture the critical feature of faster TT propagation compared to twin thickening [34, 45-48] and isolate the importance of disconnections in this process, but also allow for the development of a general model for twin embryo growth. A quantitative description of the evolution of twin length and thickness allows us to connect MD models on the nanoscale to experimental work that usually reports on a much larger length-scale. To this end, we use our atomistic observations to develop a phenomenological model describing the growth of a single twin embryo in an infinitely large grain under a constant applied shear strain. An atomic snapshot of the twin embryo is shown in Fig. 12(a), while a schematic of the phenomenological model is presented in Fig. 12(b). The coordinate system is set up the same as our atomistic simulations. In the atomic snapshot, the twin embryo is identified by lattice orientation and colored in green, being bounded by horizontal TBs, TTs, and BP/PB interfaces. This information is reproduced in the schematic, in which a green twin embryo is displayed at the center of red matrix with a length, $l$, and thickness, $h$, at a given moment in time, $t$. The TTs move in the Y-direction with velocities of $v_{TT}$, while the TBs move in the Z-direction with velocities of $v_{TB}$. Adopting the concept of BP/PB interfaces as the sources for twinning disconnections, these features are represented by the four corners of the twin embryo and generate disconnections running along the TBs. A twinning disconnection on the upper TB is identified in the atomic snapshot in Fig. 12(a), with a magnified view shown in the inset. In the schematic in Fig. 12(b), two twinning disconnections with step-heights of $h_d$ are shown on the upper and lower TBs, and these disconnections propagate along the TB at velocities of $v_d$.

The analysis of our MD results showed that the nucleation of disconnections is much faster than their propagation. For example, Figs. 5(a) and 8(c) show fast accumulation of twinning disconnections on horizontal twin planes, where new disconnections can be formed on the top of the old ones even before the old twinning disconnections managed to fully traverse across the boundary. As such, we can state that ***the propagation of twinning disconnections along the twin planes is the rate-limiting mechanism behind TB motion***. There is a connection between the twin thickening rate and its size, since the twinning disconnections move at a constant speed (Fig. 7(d)) yet the distance that needs to be traveled increases as the twin embryo grows. Homogeneous disconnection nucleation requires higher activation energy than disconnection propagation on flat TBs [28, 29], but the nucleation in our case is heterogeneous and the activation energy alone does not determine the rate-limiting process for TB migration. The distance that twinning disconnections need to travel and the time they spend travelling during TB migration is also important. Through some combination of easier nucleation of twinning disconnections from PB/BP interfaces and the time needed to traverse the TB, disconnection propagation ends up being the rate-limiting process for TB migration. In addition, our MD results show that the BP/PB interfaces are also sources of the twinning disconnections that move along the conjugate (vertical) TBs, leading to motion of the TTs. However, in this case, we do not observe any accumulation of such disconnections along the vertical boundaries, suggesting that TT motion may be rate-limited by disconnection nucleation.

The velocity of the TT can be approximated as:

$$v_{TT}(t) = \frac{h_d}{t_{tot}} = \frac{h_d}{t_{nucl}+t_{prop}} \approx \frac{h_d}{t_{nucl}} \quad (2)$$

Where $t_{tot}$ is the amount of time needed to traverse the TT. This time can be obtained as $t_{tot} = t_{nucl} + t_{prop}$, where $t_{nucl}$ is the time needed for disconnection nucleation on the TT and $t_{prop}$ is the time needed for the disconnections to propagate until they meet each other and the TT migrates one step further. $t_{nucl}$ is expected to be much larger than $t_{prop}$ according to the analysis of our MD data. In the case of the presence of both homogeneous and heterogeneous nucleation, the exact nucleation time can be determined as:

$$t_{nucl} = \min(t_{nucl}^{hom}, t_{nucl}^{het}) \quad (3)$$

where $t_{nucl}^{hom}$ is the time needed for homogeneous nucleation and $t_{nucl}^{het}$ is the time needed for heterogeneous nucleation. While it is commonly assumed that heterogeneous nucleation has a lower nucleation barrier and consequently has a shorter waiting time, this may not be true in the case of the twin embryo. The stress around the twin embryo is not homogeneously distributed, with the positive stresses found near the center of the TT, which may promote homogeneous nucleation (see the Fig. S4 of the Supplementary Material). It is also shown that the local distribution of shear stress near the corners of the twin embryo and the centers of the TTs stays relatively constant over time.

The magnitude of TT velocities is time-independent, with $v_{TT}^{right} = -v_{TT}^{left} = v_{TT} = const$. Taking an initial relaxation of the system into account, we consider that disconnection-driven steady-state twin embryo evolution begins at time $t = t_0$. Thus, the positions for the right and left TTs projected on the Y-axis are:

$$y_{TT}^{right}(t) = y_{TT}^{right}(t_0) + v_{TT}(t - t_0) \quad (4)$$

$$y_{TT}^{left}(t) = y_{TT}^{left}(t_0) - v_{TT}(t - t_0) \tag{5}$$

where $v_{TT}$ is positive for twin embryo expansion and negative for twin embryo shrinkage along the Y-axis. Thus, the expression for the twin embryo length $l = v_{TT}^{right} - v_{TT}^{left}$ is:

$$l(t) = l_0 + 2v_{TT}(t - t_0) \tag{6}$$

where $l_0 = y_{TT}^{right}(t_0) - y_{TT}^{left}(t_0)$.

Similarly, the velocities of TBs are opposite in direction and have the same magnitude, but are not a constant in time, with $v_{TB}^{upper} = -v_{TB}^{lower} = v_{TB} \neq const$. In fact, the TB migration slows down with time, as was observed in Fig. 5(b). As the twin length increases, the disconnections are nucleated from the BP/PB interfaces and moving with a constant velocity $v_d$ (see Fig. 7(d)) require more time to traverse across to push the TB position one step further. Basically, the TB velocity can be approximated as:

$$v_{TB}(t) \cong \frac{h_d}{t'_{tot}} \tag{7}$$

Where $t'_{tot}$ is the amount of time needed to traverse the TB. $t'_{tot} = t'_{nucl} + t'_{prop}$, where $t'_{nucl}$ is the time needed to emit a couple of disconnections at opposite sides of the twin embryo, and $t'_{prop}$ is the time needed for the disconnections to propagate until they meet each other and the TB migrate one step further. Due to the presence of the disconnection sources (BP/PB interfaces), the disconnection nucleation time is much smaller than the propagation time, meaning that $t'_{tot} \approx t'_{prop}$. According to the schematic in Fig. 12(b), the propagation time starting from any moment of time $t$ can be estimated as $t'_{prop} = l(t)/(2v_d)$, meaning that each of the two disconnections will on average have to traverse one half of the twin length to complete the corresponding growth of one layer. Thus, the TB velocity can be written as:

$$v_{TB}(t) = \frac{2v_d \cdot h_d}{l(t)} \tag{8}$$

The average TB position, $z_{TB}$, is then:

$$z_{TB}(t) = z_{TB}(t_0) + \int_{t_0}^{t} v_{TB}(t)dt \qquad (9)$$

By integrating the TB velocity using Eqns. (6) and (8), we can obtain expressions for the positions $z_{TB}^{upper}$ and $z_{TB}^{lower}$ of the upper and lower TBs, respectively, and, thus, obtain an expression for the twin embryo thickness, $h = z_{TB}^{upper} - z_{TB}^{lower}$:

$$h(t) = h_0 + \frac{2v_d \cdot h_d}{v_{TT}} \cdot \ln\left[1 + \frac{2v_{TT} \cdot (t-t_0)}{l_0}\right] \qquad (10)$$

where $h_0 = z_{TB}^{upper}(t_0) - z_{TB}^{lower}(t_0)$. Using Eqns. (6) and (10), the dependence of the twin embryo thickness on the twin embryo length can be also derived:

$$h = h_0 + \frac{2v_d \cdot h_d}{v_{TT}} \cdot \ln\left[\frac{l}{l_0}\right] \qquad (11)$$

Eqn. (11) can therefore be used to estimate the twin thickness at certain twin length. This phenomenological model can be modified to include the effect of homogeneous nucleation of twinning disconnections on TB migration (Supplementary Material), but adding this term gives the model an additional fitting parameter without improving the fit to the MD data for the situations targeted in this work. The aspect ratio of the twin embryo, which describes its morphology, can also be calculated as the length divided by thickness:

$$Aspect\ ratio = \frac{l(t)}{h(t)} \qquad (12)$$

To connect the phenomenological model to our MD simulations, critical parameters are extracted from MD simulations under different applied shear strains using Eqns. (6) and (10). The initial twin embryo length and thickness for MD simulations are used as $l_0$ and $h_0$ when Eqns. (6) and (10) are fitted to the MD data, and the fitting results are shown in Fig. 13. The fit is excellent for all of the MD simulations performed in this work. Small deviations from a linear relationship between twin embryo length and time are observed at the very early times for samples deformed by 8% to 10% shear strain, due to an early twin embryo morphology that deviates from the idealized rectangular form (see inset of Fig. 13(a) at 5 ps). The twin embryo that grows at 10% shear strain initially adopts a more elliptical morphology that is rotated with respect to the simulation cell axes, relaxing some of the assumptions made in the model. This deviation becomes less notable as the twin embryo grows (see inset of Fig. 10(a) at 25 ps). TBs and TTs are slightly tilted because of the disconnection accumulation at the top right and bottom left corners.

Fig. 14(a) shows the twin aspect ratio at different times during the MD-simulated twin embryo growth process for the sample deformed at 7% shear strain. The phenomenological model captures the important fact that the twin embryo aspect ratio first drops, and then quickly increases, although it misses a few minor details from the MD model. Fig. 14(b) displays the velocities of the TBs and TTs for the sample deformed at 7% shear strain. While the velocity of the TT remains constant over time, the TB velocity decreases as the twin embryo grows. While the TBs initially grow faster than the TTs, given the initial drop in aspect ratio, the TB velocity quickly falls below the constant TT velocity.

To explore the influence of shear stress on TB and TT motion, values for $v_d$ and $v_{TT}$, two key velocities that describe twin embryo growth, are extracted for different matrix shear stress values. The matrix shear stress is calculated by straining a sample without a twin to different strain levels (shown in Fig. S1). As shown in Fig. 15(a), both velocities increase with the matrix shear stress, which can be expected, since the applied shear stress is the driving force for disconnection migration and twin embryo growth. Also important for determining twin morphology is the ratio of $v_d$ to $v_{TT}$ as a function of matrix shear stress. Fig. 15(b) shows that this ratio also monotonically increases with the matrix shear stress. We should mention that the same result is obtained for initial twin embryos of different shapes and sizes. As seen in Fig. 15(b) the dashed blue line obtained by fitting all the data points nearly overlaps the dashed black line obtained by only fitting the data points of samples with embryos having $l_0/h_0 \approx 1.6$, indicating that the initial twin embryo shape and size does not alter the relationship between the velocity ratio and the matrix shear stress. In the following discussion, we will only use the data for embryo with $l_0/h_0 \approx 1.6$. The variation of twin length/thickness with time for initial embryos having $l_0/h_0 \approx 3.7$ and $l_0/h_0 \approx 1.1$ is shown in Fig. S8 of the Supplementary Material, together with the fitting curves. Finally, we note that the disconnection velocities in Fig. 15 are larger than the values shown in Fig. 7(d) from the MD simulations. The main reason behind this discrepancy is that homogeneous nucleation of twinning disconnections does occasionally occur, which is not included in our model that only considers disconnection nucleation from the BP/PB interfaces. With slightly more disconnections than our model predicts, each needs to move slower to give the same twin thickening rate. Even with a number of approximations, the simple phenomenological model can accurately recreate the important phenomenological features of twin embryo growth.

The experimental work of Yu et al. [34] with in situ tensile, compressive, and bending tests on Mg single crystals of nanometer scale was chosen to test our phenomenological model. This work was chosen because: (1) it is known that the twin embryos nucleate at free surfaces, (2) no grain boundaries exist surrounding the twin, and (3) the tensile samples are nanometer sized and therefore dislocation-free. These factors result in a relatively uniform stress field, without abrupt local stress concentrations associated with other microstructural features. The experimental setup is schematically shown in Fig. 16(a), and a bright field transmission electron microscopy image showing an array of $\{10\bar{1}2\}$ twins is displayed in Fig. 16(b). The widths of the tensile bars were reported by the authors to be in the range of 100-200 nm, with the example shown in Fig. 16(b) being 100 nm wide. Forces were applied along the c-axis, and the nucleation of the $\{10\bar{1}2\}$ twins started at ~ 800 MPa. The twins then elongated during testing, with some of the TTs reaching the opposite surface of the sample. Most of the twins have thicknesses of 5-10 nm. To be more consistent with the experimental setup, the phenomenological twin embryo growth model was modified to include the existence of a free surface or a grain boundary that can provide sites for heterogeneous nucleation of twinning disconnections [34, 49]. Free surfaces and grain boundaries should have a similar ability to generate disconnections as BP/PB interfaces. Most of these twins observed in the experiment adopt a half-lenticular morphology, meaning that they are thinner at the TTs terminated in the matrix and wider at the free surface or grain boundary [34, 50, 51].

The schematic for the modified twin embryo growth model is presented in Fig. 16(c). The TT cannot move to the left due to the presence of the free surface, so Eqns. (6), (10) and (11) are updated as:

$$l(t) = l_0 + v_{TT}(t - t_0) \tag{13}$$

$$h(t) = h_0 + \frac{4v_d \cdot h_d}{v_{TT}} \cdot \ln\left[1 + \frac{v_{TT} \cdot (t-t_0)}{l_0}\right] \tag{14}$$

$$h = h_0 + \frac{4v_d \cdot h_d}{v_{TT}} \cdot \ln\left[\frac{l}{l_0}\right] \tag{15}$$

Eqn. (15) can be used to fit experimental data of twin length and thickness to obtain the ratio of disconnection velocity to TT velocity. Measurements of twin length and twin thickness from the image displayed in Fig. 16(b) are presented in Fig. 16(d). Nine twins that could be clearly identified were selected. The twin length and twin thickness is then approximated as the length and width of the rectangles shown in Fig. 14(b), respectively. The ratio of $v_d/v_{TT} = 3.4 \pm 0.55$ is obtained. To determine the shear stress, the applied stress at the point of twin nucleation that was reported in experiments is transformed to the resolved shear stress ($\tau_{RSS}$) via the Schmid factor ($m$).

The $\tau_{RSS} = \sigma \cdot m = \sigma \cdot (\cos\theta \cdot \cos\varphi)$, where $\theta$ is the angle between the loading axis and twin plane normal, $\varphi$ is the angle between the loading axis and the twinning direction, and $\sigma$ is the applied normal stress. For the $\{10\bar{1}2\}$ twins created by stretching the c-axis, the Schmid factor is 0.499, so $\tau_{RSS}$ = 800 MPa × 0.499 = 399.2 MPa. Fig. 16(e) shows this single experimental data point together with our MD data. The experimental data point fits well with the extrapolated curve for our MD simulations at 300 K. The fact that the ratio of the disconnection velocity to TT velocity can be predicted and is consistent with our atomistic modeling results demonstrates the robust nature of the phenomenological model developed here. The heterogeneous nucleation of twinning disconnections dominates TB migration at twin lengths of several hundred nanometers. Note that as the matrix shear stress decreases to a certain level, the velocity ratio $v_d/v_{TT}$ becomes negative, indicating the shrinkage of an existing twin embryo. This finding is again consistent with our MD simulations, which showed that a critical strain must be applied or else the twin embryos will shrink and disappear. Eqn. (15) also yields a relation between the twin length ($l_0$) and twin thickness ($h_0$) at the beginning of the steady-state twin embryo evolution, which is $5.168 \cdot \ln l_0 - h_0 = 17.96$. So if a range of $h_0$ from 0.76 nm (twice the step height of twinning disconnections and the minimum imaginable twin thickness) to 4.3 nm (initial twin thickness used in MD simulations) was considered, then the expected range for $l_0$ would be 37 nm to 74 nm. We note that experiments typically are unable to extract the values of the initial twin embryo size, as it would necessitate high magnification transmission electron microscopy (TEM) in exactly the right place at exactly the right time.

The current model is developed to describe the initial embryotic growth stage of a twin formed by homogeneous nucleation in an infinitely large and perfect grain, in which there are no impurities or solute atoms, defects such as dislocations, and influence from other twins. The nucleation process of twin embryos, such as where and how twin embryos are nucleated are not included in this work. The model can be applied to analyze experimental results, especially the relationship between twin length and twin thickness described by Eqn. (15) can be used to predict the final thickness of twins. Parameters, such as the initial size of the twin embryo, $l_0$ and $h_0$, the final twin length, $l$, and the ratio of the two velocities, $v_d$ and $v_{TT}$ must be obtained. While the final twin length could be measured via electron microscopy and the ratio of the velocities can be predicted by MD simulations for a given set of conditions, the initial size of the twin embryo is less certain. As mentioned earlier, experimental observations of twin embryos are missing due to their rapid expansion, with most twin embryos growing beyond their initial size by the time of observation. MD simulations allow twin nucleation to be explored. For instance, twin embryos can be initiated by structurally relaxing two partial dislocations at a certain distance [22], meaning the initial twin embryo size might also be found from such simulations. Our findings are also applicable to the thickening of "tapering twins". Mahajan and coworkers [52-54] first observed twinning dislocations and their reactions at thin "tapering twins" in face-centered-cubic Cu and body-centered-cubic Mo-Re alloys. "Tapering twins" may be formed by tiny embryo growing into each other, and they contain twinning dislocations/disconnections. These disconnections could move in a similar fashion.

A possible limitation of our simulations and phenomenological model is the two-dimensional nature. Recently, Liu et al. [21] revealed the structure of the "dark side" of a twin embryo via high resolution transmission electron microscopy. They found the "dark side" of a twin embryo is composed of coherent TBs and semi-coherent twist prismatic–prismatic boundaries. To explore the migration of the "dark side", Gong et al. [55] developed the 3D twin embryo in MD simulations, and reported the motion of twist prismatic–prismatic boundaries through atomic shuffling as well as the pinning effect of misfit dislocations on such boundaries. Spearot et al. [27] also showed that three-dimensional simulations are needed to capture disconnection terrace nucleation and growth processes that are essential for the migration of mature TBs (i.e., larger TBs which can be approximated as flat interfaces, without considering bounding PB/BP interfaces, grain boundaries, or surfaces). In this study, the third dimension was essential for capturing homogeneous disconnection nucleation from the flat TB. In our paper, we focus on the growth of an embryo in its shear plane in two-dimensions, so any interactions with defects responsible for growth in the third dimension are not considered here. However, we obtain excellent agreement with experiments [34], which would indicate that interactions with processes in the third direction do not dominate the early stages of embryo growth.

In addition, the current model can be extended in multiple ways to cover different conditions in experiments, which opens up future research opportunities. As discussed earlier, grain boundaries and free surfaces may also facilitate the nucleation of twinning disconnections. Twinning disconnections might also be generated through the interaction between twins and dislocations [13]. For the case where multiple twins grow in one grain, a common occurrence observed in experiments, the stress fields around the twins can interact and affect their subsequent growth. For example, using tri-crystal models containing two twins, Arul Kumar et al. [56] predicted that within a certain distance, the stress-fields around two twins overlap and give higher back stresses that resist further twin growth. As twins get thicker and longer, it is also possible that multiple twins merge together and eventually leave one thick twin in the grain, or twins propagating in different directions might intersect to form twin junctions. These junctions can also play an important role in twin growth, as demonstrated in early studies on Co and Co-Fe showing that the incident shear strains of the crossing twin can be accommodated by activating dislocation slip within or on the exit side of the crossed twin, or alternatively inducing secondary twinning or secondary slip in the crossed twin which can result in either thickening or detwinning [57, 58]. Segregating impurities or solutes are another contributing factor that strongly influence twin growth. Solutes located at the boundary of the twin embryo demonstrate a pinning effect on boundary motion and thus suppress twin growth, as shown in the work of Nie et al. [51]. These authors used atomic resolution transmission electron microscopy images of Gd-doped and Zn-doped coherent TBs in a Mg alloy to show that solute segregation limited the expansion of twins, compared to observations in pure Mg.

## 5. Conclusions

In this work, MD simulations were conducted to reveal the initial growth mechanisms of a twin embryo in a single Mg crystal. The simulation results reveal a constant velocity for the TT, while a decrease in the TB velocity over time (i.e., as the twin embryo grows). Twinning disconnections nucleated at the intersections between BP/PB interfaces and the TBs, followed by their lateral motion across the TB. The velocity of the twinning disconnections was found to be constant for a given applied shear stress. Accumulation of twinning disconnections on TBs occurred, especially at the later stage of simulation when the twin length is relatively large, suggesting that the propagation of twinning disconnections is the rate-limiting process for twin embryo growth.

In addition, a phenomenological model was developed to describe the evolution of twin length and thickness with time. The proposed model fits well with the simulation data for the two key parameters, the velocities of TTs and twinning disconnections, being extracted. Both velocities increase monotonically as the applied shear stress increases. The phenomenological model was also found to fit an experimental report on nanoscale twin growth, even though the stress level was much lower than those probed by MD and used to calibrate the model. We envision that this model can be used to predict various twin geometries and morphologies given known experimental conditions.

Taken as a whole, our work fills an important gap in the understanding of twinning by providing a clear and consistent description of the early stages of twin embryo growth. The time evolution of twin length and thickness are studied together for the first time, with kinetic parameters such as TT velocity and twinning disconnection velocity obtained.

**Acknowledgments**

The authors acknowledge financial support from the National Science Foundation through grants CMMI-1729829, CMMI-1729887, and CMMI-1723539.

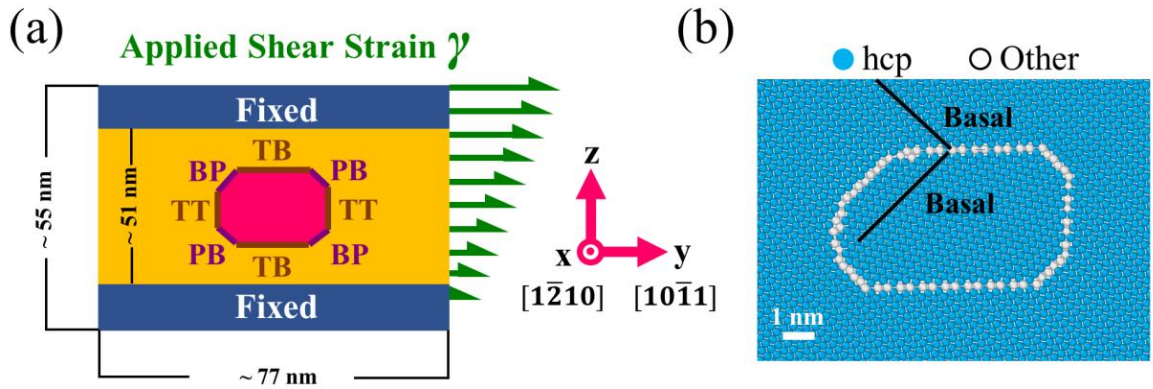

Fig. 1. (a) A schematic of the shear deformation simulation, and (b) an atomic snapshot of the initial twin embryo. In (b), hcp atoms are colored blue, while atoms at the boundaries are colored white due to their local structure. The different interfaces associated with the twin embryo are designated in (a). Shear strains are applied by displacing all atoms in the simulation cell, as it is shown by the green arrows in (a) which denote the displacement at a given height within the sample. The basal planes in the matrix and the twin are shown in (b) using solid black lines.

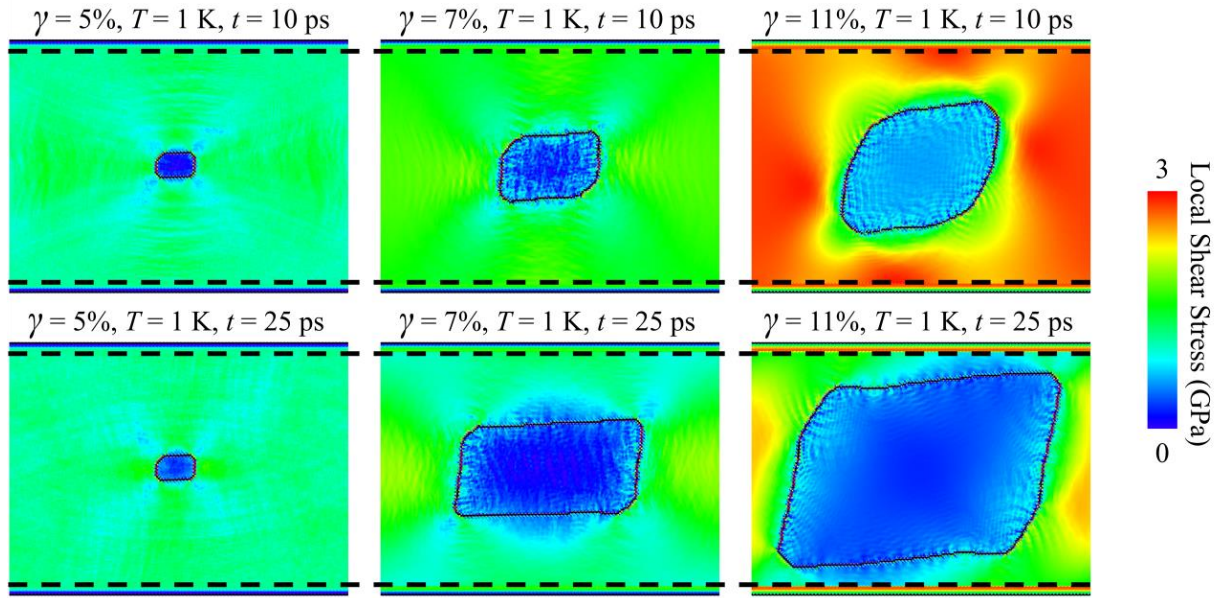

Fig. 2. The distribution in atomic shear stress from snapshots taken at 10 ps and 25 ps for samples deformed at 5%, 7% and 11% shear strain and 1 K. The mobile-atom region and fixed atom region are separated by black dashed lines, while the boundaries of the twin embryo are denoted by a solid black outline.

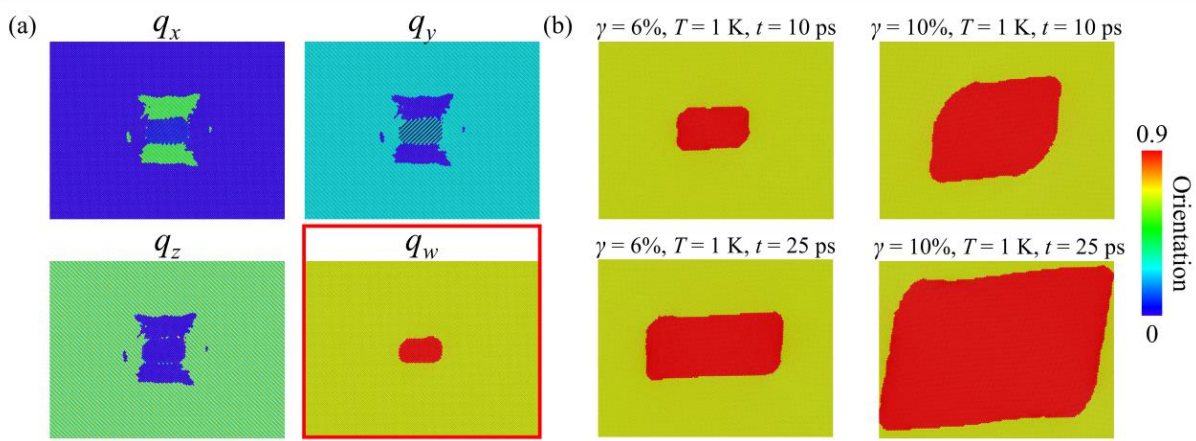

Fig. 3. (a) The local orientation information associated with each atom in the MD samples obtained by Polyhedral Template Matching. The twinned region can be correctly selected by $q_w$ with a value larger than 0.8, and this criterion is later used in determining the boundaries of the twin embryo. (b) The twinned region selected by the criterion, $q_w > 0.8$, in samples deformed at 6% and 10% shear strains and at 10 ps and 25 ps, respectively.

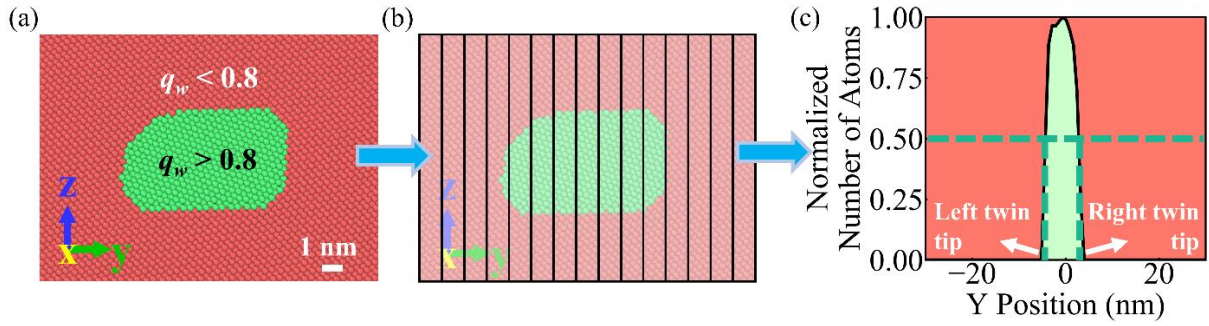

Fig. 4. Determining the positions of twin embryo boundaries. (a) The twinned region can be accurately detected using lattice orientation information. The twin embryo is colored green to contrast with the red matrix. (b) The simulation box is then divided into several bins, and the number of atoms with $q_w$ larger than 0.8 in each bin is counted. (c) Next, these numbers are normalized by their maximum value and plotted versus the positions of the bins, where the peak represents the twinned region while the locations of boundaries are chosen as the two positions with Y-axis value equal to 0.5.

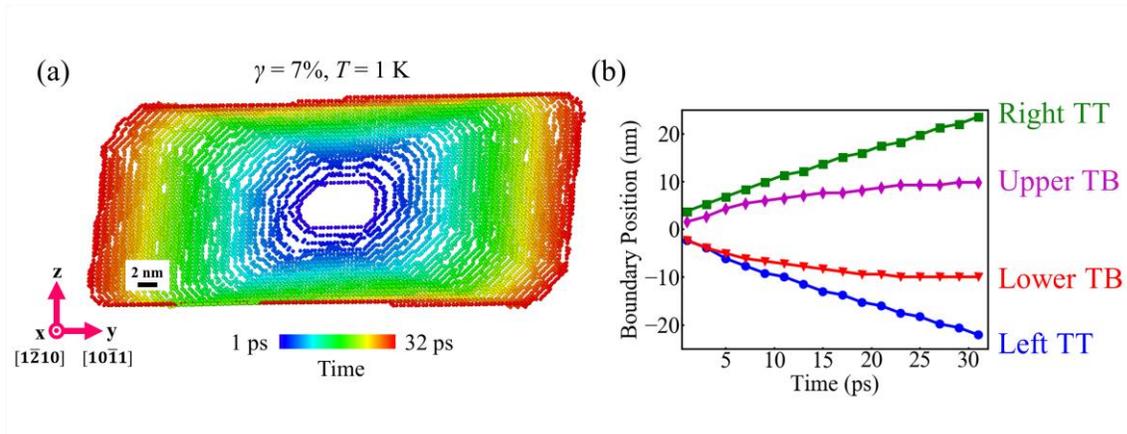

Fig. 5. Time evolution of the twin embryo at 7% shear strain and 1 K. (a) Boundary atoms of the twin embryo are colored by the time at which they are extracted. (b) The positions of boundaries versus time.

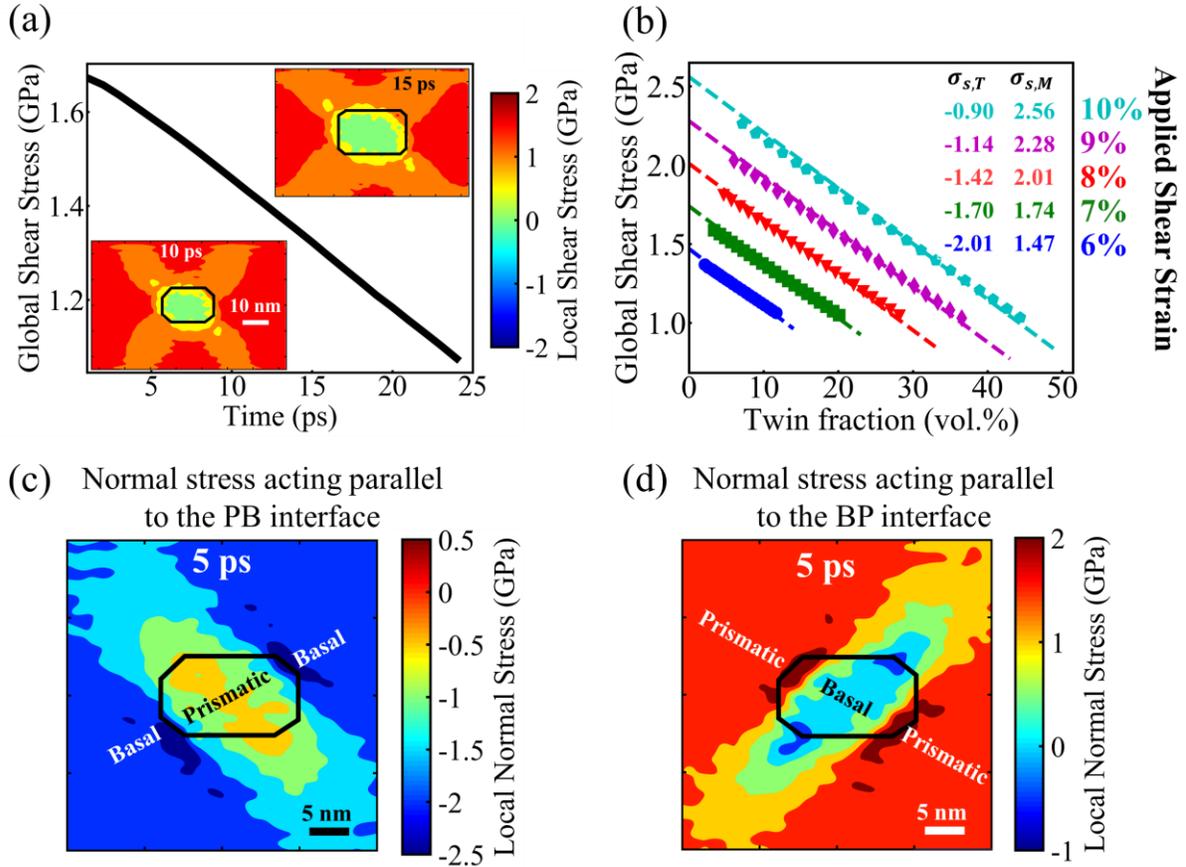

Fig. 6. (a) The variation in global shear stress with time for the samples deformed at 7% shear strain and 1 K. The two insets show the shear stress field around the twin embryo at 1 ps and 25 ps, respectively. (b) The variation in global shear stress with the twin volume fractions for samples deformed at different shear strains. The inset shows the contribution of matrix and twin to the global shear stress. (c) The spatial distribution of normal stresses acting parallel to the PB interface at 5 ps. (d) The spatial distribution of normal stresses acting parallel to the BP interface at 5 ps.

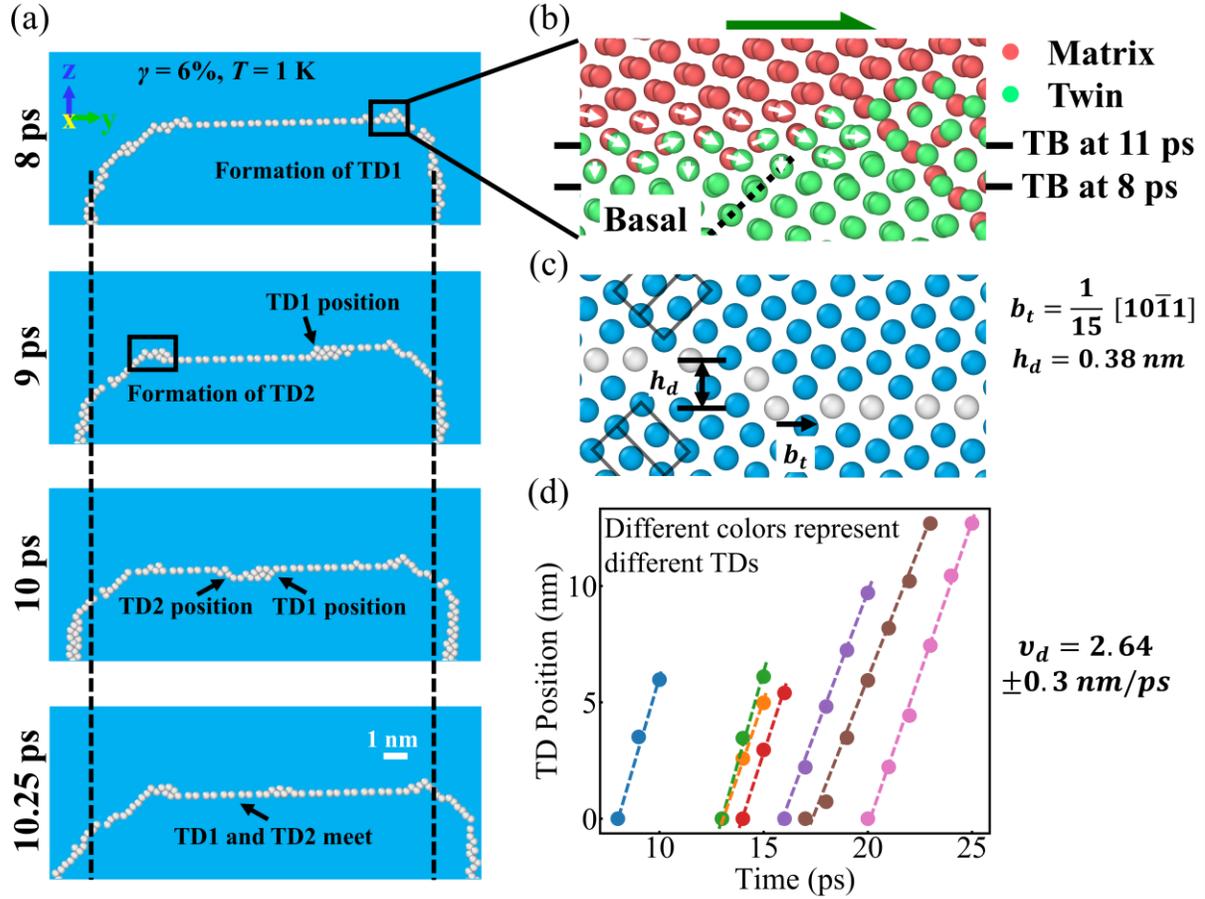

Fig. 7. (a) The nucleation and propagation of twinning disconnections on the upper TB in a sample deformed at 6% shear strain and 1 K. The positions of two twinning disconnections (TD1 and TD2) are marked in each frame. The dashed black lines represent the positions of TTs at 8 ps. (b) The spatial distribution of atoms near TD1 at 8 ps and 11 ps, where atoms in the matrix are colored red, while those in the twin embryo are colored green. The initial and new TBs are marked using solid black lines, while the dashed black line shows the basal plane in the twin. The white arrows indicate the atomic displacements. The dark green arrow on the top of the figure shows the direction of the applied shear strain. (c) The zoomed-in view of a twinning disconnection with its Burgers vector ($b_t$) and step character ($h_d$) shown, the outlines of two hcp cells are also drawn in the matrix and twin. The black arrow shows the Burgers vector of the twinning disconnection, the magnitude of which is very small. (d) The variation in disconnection position versus time. Data points in different colors are obtained for different twinning disconnections.

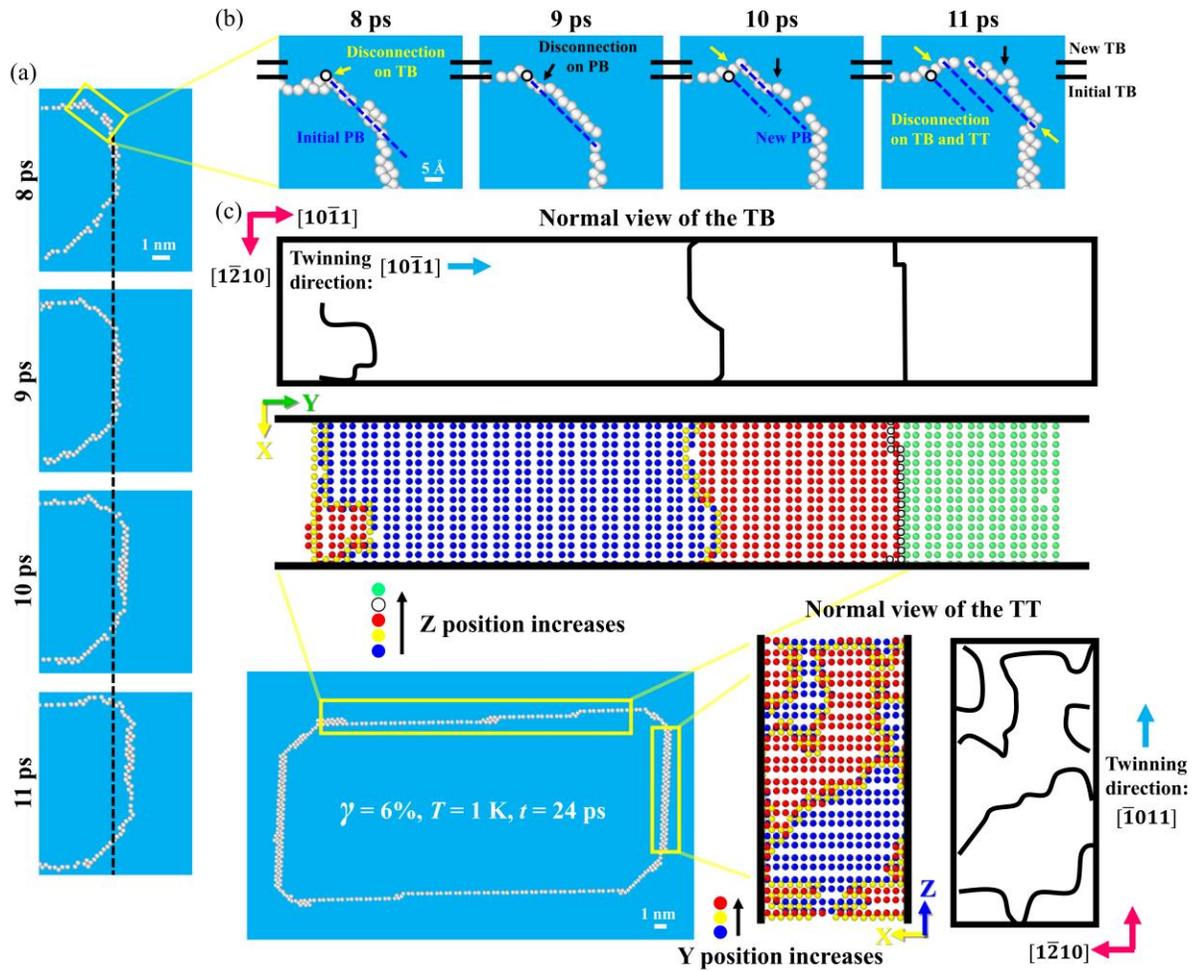

Fig. 8. (a) The atomic snapshots of the right TT from 8 to 11 ps for the sample deformed at 6% shear strain and 1 K. (b) The magnified view of the PB interface from 8 to 11 ps. (c) A comparison of the twinning disconnections on TB, as well as TT; atoms are colored by their Y positions or Z positions. In (a) the TT position at 8 ps is marked using a black dashed line. In (b), the TB positions at 9 ps and 11 ps are marked using black solid lines, while the PB interface positions at 8 ps, 10 ps and 11 ps are marked using dark blue dashed lines. Disconnections formed on the TB and the TT are marked by yellow arrows, while disconnections formed on the PB interface are marked by black arrows. The atom circled in black is a reference atom; its position does not change in all the frames. In (c), disconnection lines are marked using black lines, and blue arrows show the twinning direction for each plane.

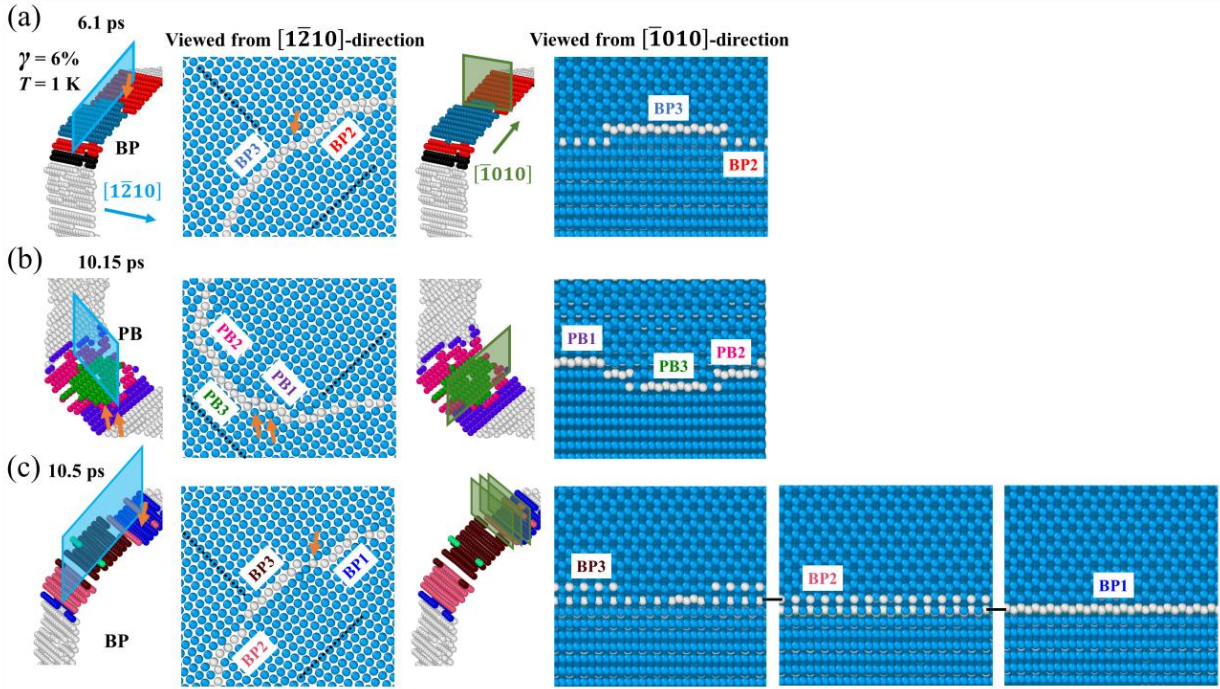

Fig. 9. Disconnections formed on BP/PB interfaces at (a) 6.1 ps, (b) 10.15 ps, and (c) 10.5 ps. Samples are deformed at 6% shear strain and 1 K. Different BP/PB planes are colored differently, in order to be identified more clearly, and ordered according to the time sequence of formation, with smaller numbers denoting earlier formation. The atomic structures of the disconnections are viewed from the $[1\bar{2}10]$-direction (a direction perpendicular to the blue parallelogram) and $[\bar{1}010]$-direction (a direction perpendicular to the green parallelogram), respectively. Black dashed lines are used to show the orientations of basal planes in the matrix and twin, while orange arrows are used to mark individual disconnections.

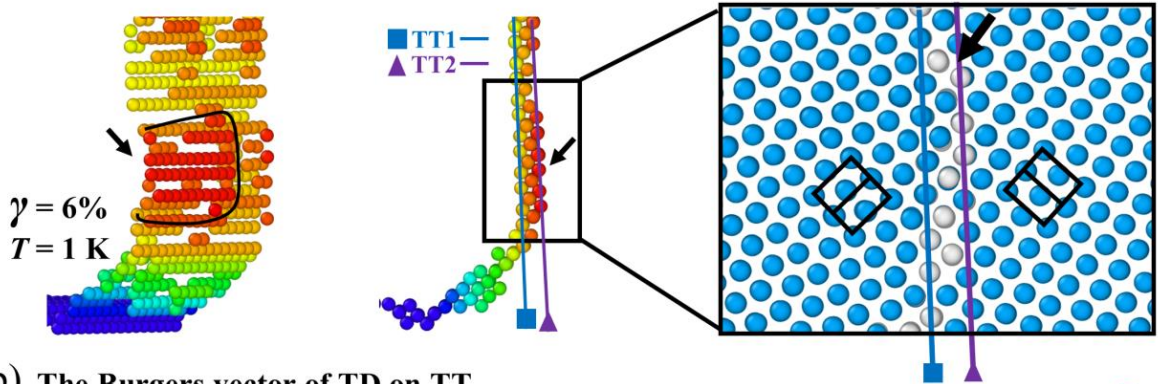
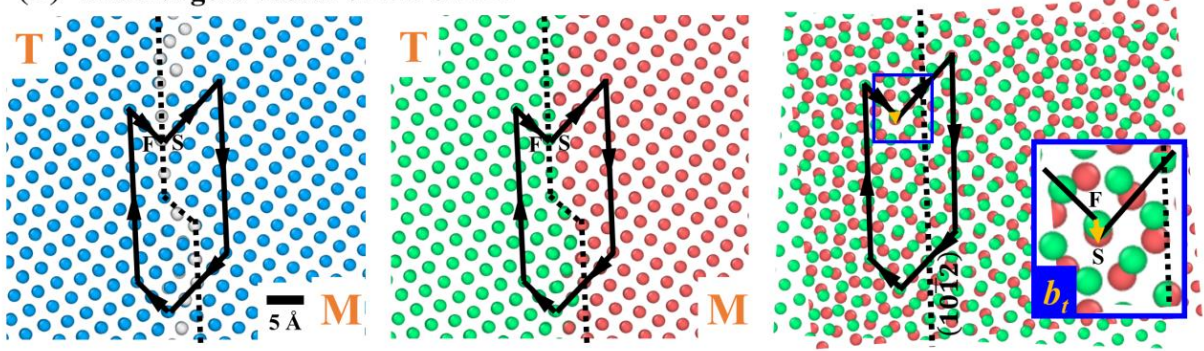

Fig. 10. (a) The atomic structure of one twinning disconnection formed on the TT in a sample deformed at 6% shear strain and 1 K, where the disconnection line is marked with black lines and the old and new positions of the TT are marked using solid blue and purple lines, respectively. Black arrows are used to show the location of the twinning disconnection. The outlines of two hcp cells are drawn in the matrix and twin. (b) Determination of the Burgers vector of the twinning disconnection shown in (a). The Burgers vector ($b_t$) of the twinning disconnection is marked using yellow arrows in the last frame, with the inset giving a magnified view of the Burgers vector. The structure is viewed from the negative X-direction. "M" and "T" signify matrix and twin, respectively.

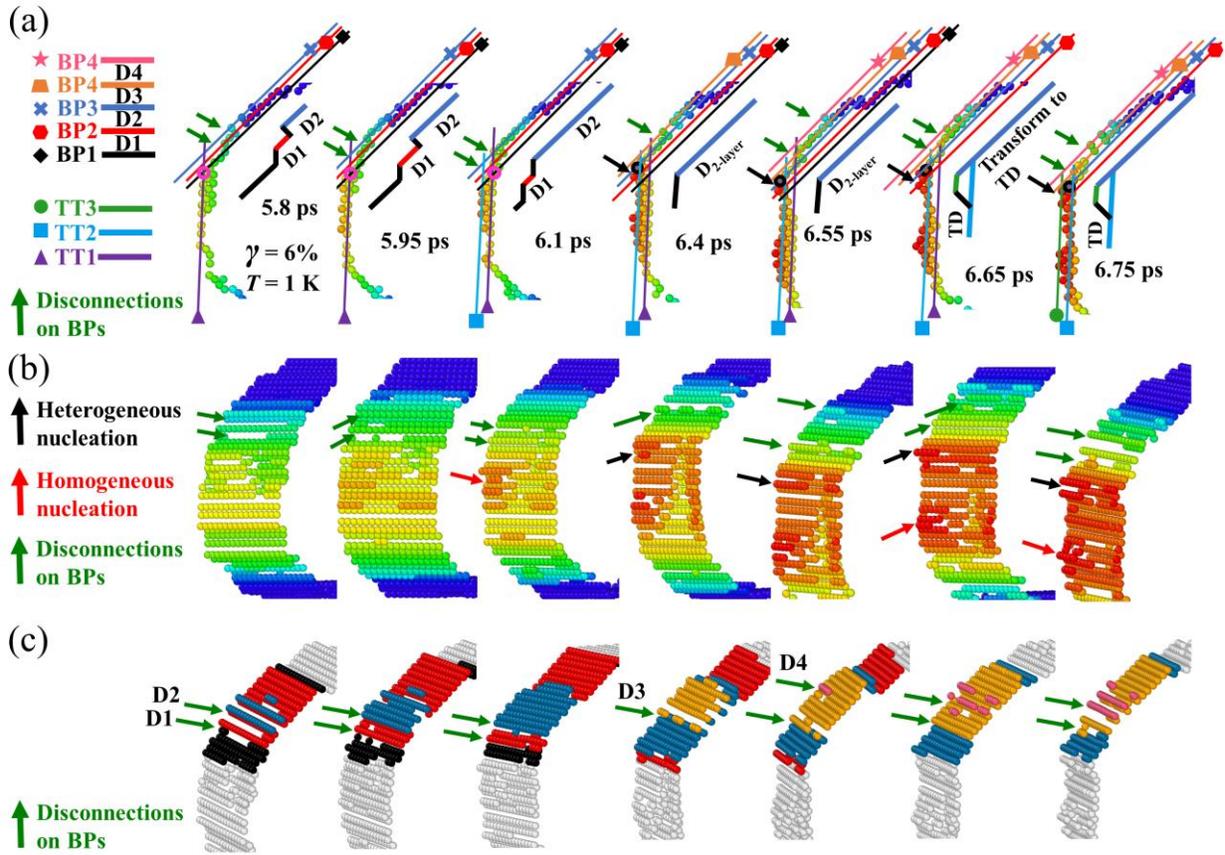

Fig. 11. Disconnections formed on the TTs and BP interfaces during the growth of the twin embryo in a sample deformed at 6% shear strain and 1 K, shown in both (a) normal and (b) perspective views. (c) Simplified view of disconnections on the BP interface. Different BP planes and TTs are colored differently and ordered according to the time sequence of formation, with smaller numbers denoting earlier formation. Black and dark green arrows are used to mark the disconnections formed on the TT and BP planes, respectively. Disconnections formed on BP planes are also ordered according to the time of formation. Red arrows mark two disconnections formed homogeneously on the TT. The atoms circled in magenta and black are reference atoms. The position of the atom circled in magenta does not change from 5.8 ps to 6.1 ps, while the position of the atom circled in black does not change from 6.4 ps to 6.75 ps.

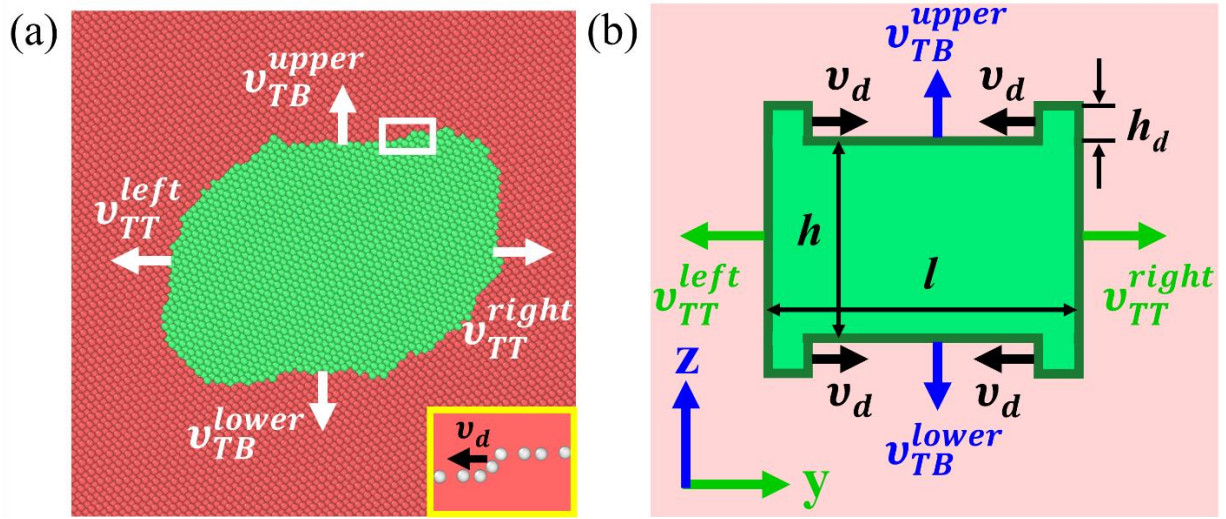

Fig. 12. (a) The atomic snapshot of a twin embryo in MD simulations with an inset of a twinning disconnection. (b) Schematic of the twin embryo growth model. The twin embryo in (b) is bounded by two TBs and two TTs, and the BP/PB interfaces as the source of the twinning disconnections are represented by the four corners.

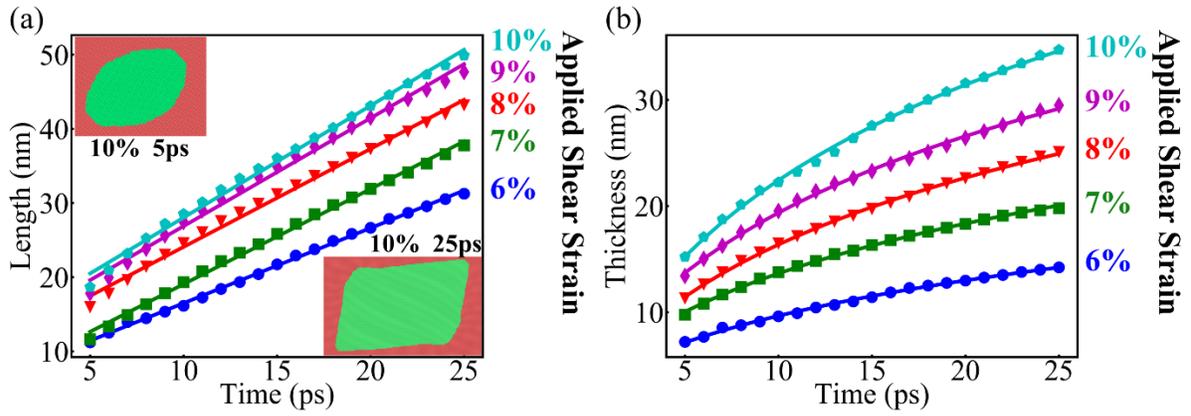

Fig. 13. Time evolution of the (a) length and (b) thickness of the twin embryo that grows at different shear strains and 1 K. The length is fitted by Eqn. (6), while the thickness is fitted by Eqn. (10). The insets in (a) show the twin embryo in a sample deformed at 10% shear strain at 5 ps and 25 ps. The twinned region is colored green, while the matrix is colored red.

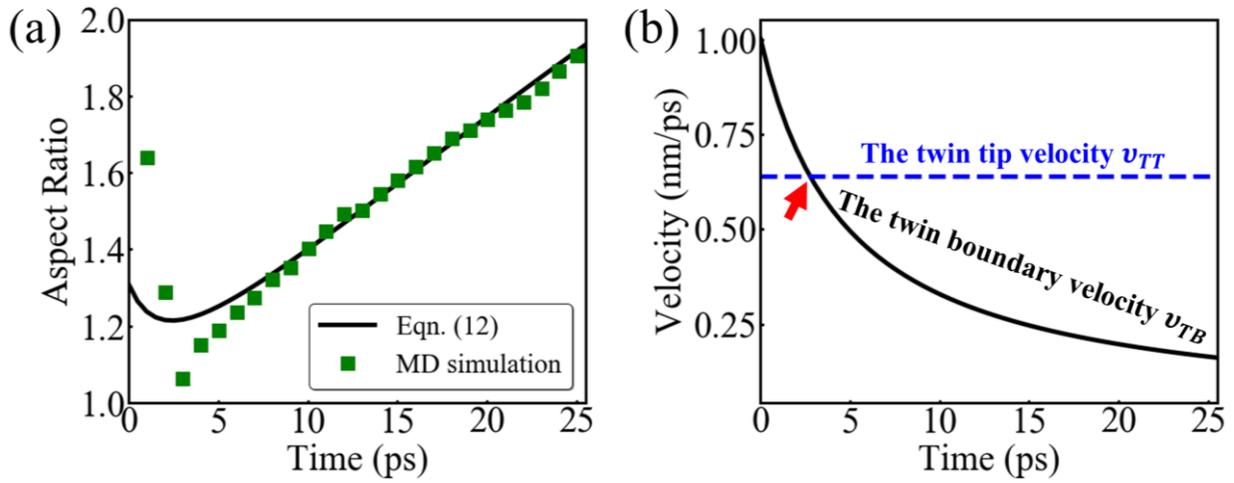

Fig. 14. (a) The aspect ratio of the twin embryo versus time for the sample deformed at 7% shear strain and 1 K. The black curve is given by Eqn. (12). (b) TT velocity and TB velocity versus time for the same sample. The red arrow shows the intersection of two curves indicating a flip in magnitude of the two velocities.

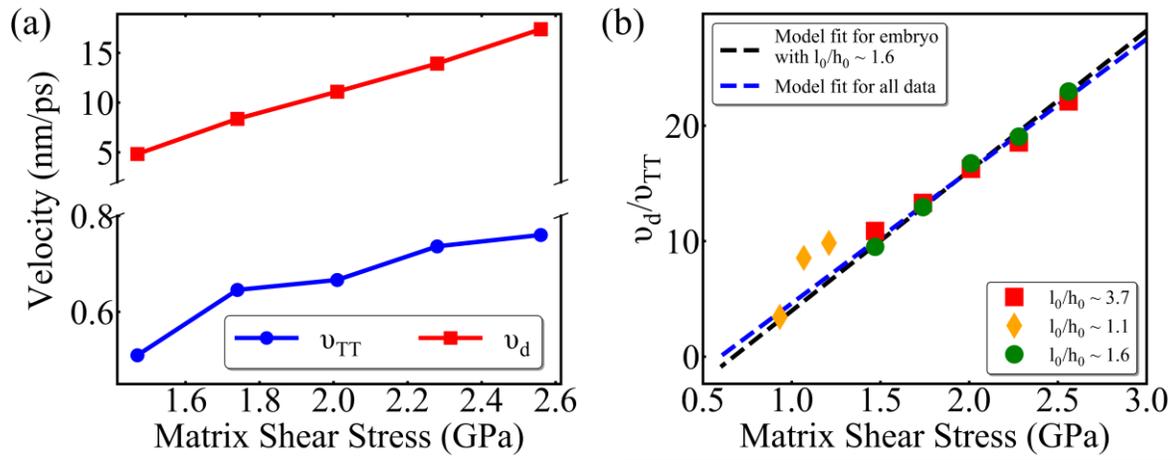

Fig. 15. (a) $v_{TT}$ and $v_d$ versus the matrix shear stress. (b) The ratio of $v_d$ to $v_{TT}$ versus the matrix shear stress for samples with different initial sizes of twin embryos.

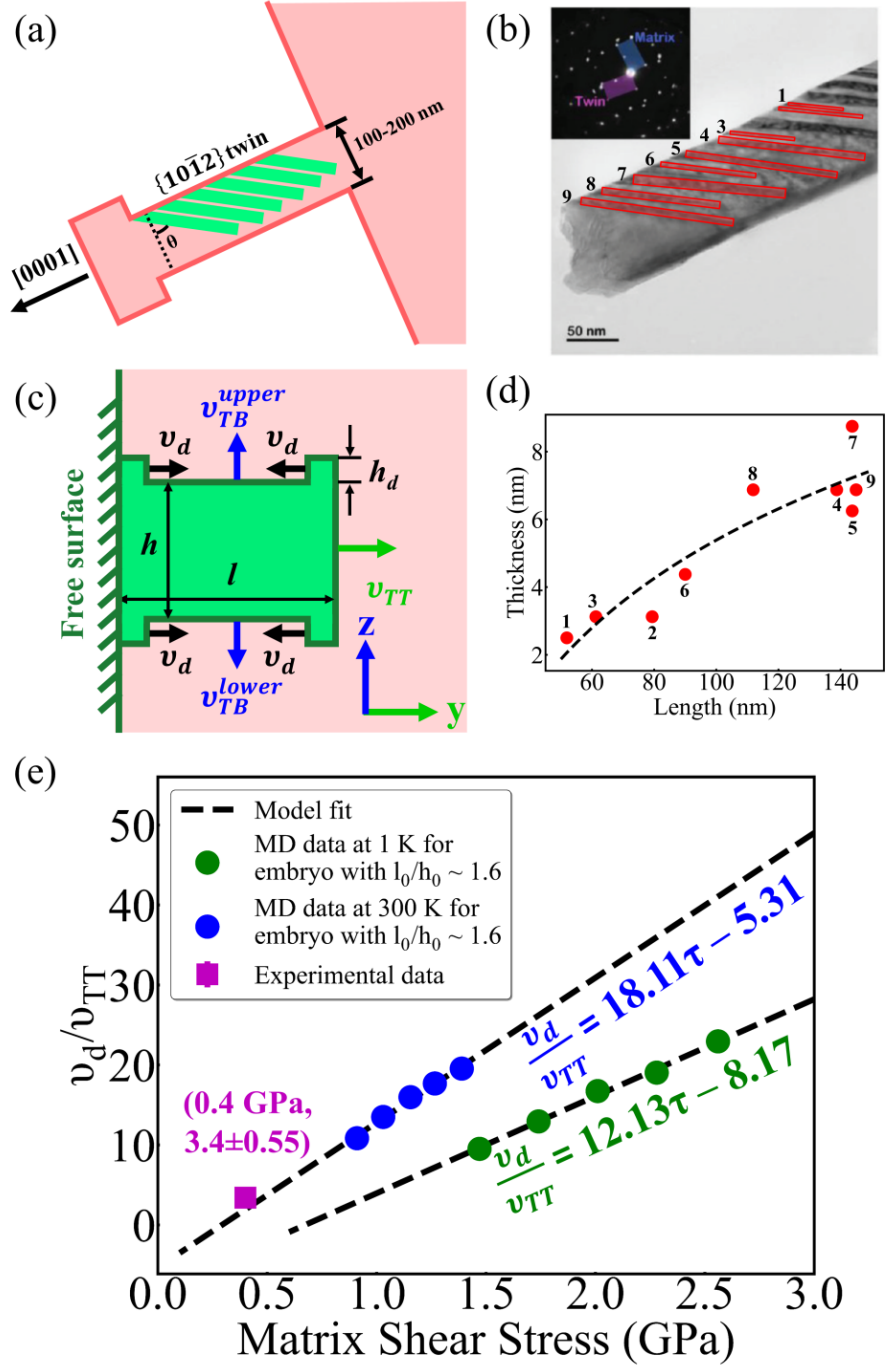

Fig. 16. (a) The experimental set up of the *in-situ* tensile testing in Ref. [34]. (b) A TEM image that shows the $\{10\bar{1}2\}$ nanotwin array in a tensile sample. The inset displays the related diffraction pattern with beam direction along $[1\bar{2}10]$. Nine twins are selected to measure length and thickness and are numbered from 1 to 9. The experimental data come from Ref. [34]. (c) The schematic of the modified twin embryo growth model with the presence of free surfaces. (d) The twin thickness versus twin length for those twins selected from (b) along with the fit obtained using Eqn. (15). (e) The ratio of $v_d$ to $v_{TT}$ versus matrix shear stress for both MD simulations and data extracted from the experimental work [34].